\documentclass[12pt]{iopart}
\usepackage{bbold}
\usepackage{graphicx}

\expandafter\let\csname equation*\endcsname\relax
\expandafter\let\csname endequation*\endcsname\relax
\usepackage{amsmath,amssymb}

\newcommand{\be}{\begin{equation}}
\newcommand{\en}{\end{equation}}

\newcommand{\bbra}[1]{\langle #1 |}
\newcommand{\ket}[1]{| #1 \rangle}

\DeclareMathOperator{\arccosh}{arccosh}

\newcommand{\Laplayer}{\Delta_{d-1}}

\newcommand{\intinf}{\int_{-\infty}^{\infty}}

\newcommand{\bx}{\boldsymbol{x}}
\newcommand{\by}{\boldsymbol{y}}
\newcommand{\bb}{\boldsymbol{b}}

\renewcommand{\exp}[1]{{\rm exp}\left[#1\right]}
\newcommand{\diag}{{\rm diag}}

\newcommand{\bpsi}{\boldsymbol{\psi}}

\newcommand{\bn}{\boldsymbol{n}}
\newcommand{\bm}{\boldsymbol{m}}

\newcommand{\dell}{\delta L}

\newcommand{\bbr}{\boldsymbol{r}}
\newcommand{\bbrho}{\boldsymbol{\rho}}

\newcommand{\prodrho}{\prod_{i=1}^{d-1}}

\begin{document}
\title{Gelfand-Yaglom formula for functional determinants in higher dimensions}
\author{A. Ossipov}
\address{School of Mathematical Sciences, University of Nottingham, Nottingham NG7 2RD, United Kingdom} 
\ead{alexander.ossipov@nottingham.ac.uk}
\begin{abstract}
The Gelfand-Yaglom formula relates functional determinants of the one-dimensional second order differential operators to the solutions of the corresponding initial value problem. In this work we generalise the Gelfand-Yaglom method by considering discrete and continuum partial second order differential operators in higher dimensions. To illustrate our main result we apply the generalised formula to the two-dimensional massive and massless discrete Laplace operators and calculate asymptotic expressions for their determinants. 
\end{abstract}
\maketitle


\section{Introduction}

Functional determinants of the second order differential operators appear naturally in many different fields  ranging from quantum mechanics \cite{Kleinert_book}, to quantum field theories \cite{D08}, to condensed matter physics \cite{Altland_book, C78, HJ91, FF15, FFG17, FDRT17} and mathematics \cite{RS71, F87}. Generally, computation of a functional determinant represents a very non-trivial problem and requires the knowledge of all eigenvalues of the differential operator. Even when all eigenvalues are known, calculation of their infinite product might be challenging. However in one-dimensional case, there is an elegant method developed by Gelfand and Yaglom (GY) \cite{GY60}, which allows one to calculate a functional determinant by solving an initial value problem for the corresponding differential operator. Consider a one-dimensional differential operator
\begin{eqnarray}
H=-\frac{d^2}{dx^2}+V(x),
\end{eqnarray}
where $V(x)$ is an arbitrary potential and the operator acts on functions $\psi(x)$ defined on the finite interval $[0,L]$ and satisfying the Dirichlet boundary condition $\psi(0)=\psi(L)=0$. Suppose that one can solve the corresponding initial value problem
\begin{eqnarray}
-y''+Vy=0,\quad y(0)=0,\; y'(0)=1.
\end{eqnarray}
Then the GY formula reads
\begin{eqnarray}
\det \left(-\frac{d^2}{d x^2}+V(x)\right)= y(L).
\end{eqnarray}
As the functional determinants usually diverge, this formula should be understood either in the discrete setting \cite{D12} or using some regularisation, for example,  by considering a ratio of two functional determinants with and without $V(x)$ \cite{D08, F87,MT95,KM03,D12}.  Thus in one dimension, functional determinants can be computed without knowing the eigenvalues explicitly.    

The aim of this work is to derive the generalisation of the GY formula to the higher dimensional case. Specifically, we relate the functional determinant of the differential operator 
\begin{eqnarray}
H &=&-\Delta_d+V(\bbr),
\end{eqnarray}
where $\Delta_d$ is the $d$-dimensional Laplacian acting on functions $\psi(x,\bbrho)$ defined on  the $d$-dimensional box, $\bbr\equiv(x,\bbrho)$, $x\in [0,L],\: \rho_k\in [0,W], \: k=1,\dots d-1$ and supplemented with the Dirichlet boundary conditions in all directions, to the matrix solution, $\hat{Y}(x)$, of the initial value problem  
\begin{eqnarray}
-\hat{Y}''+(-\hat{\Delta}_{d-1}+\hat{V})\hat{Y}=0, \quad \hat{Y}(0)=0,\;\hat{Y}'=I,
\end{eqnarray}
where $\hat{A}$ denotes a matrix of the operator $A$ w.r.t. any complete countable basis of the functions $\{u_l(\bbrho)\}$. The analogue of the GY formula is then given by
\begin{eqnarray}
\det \left(-\Delta_d+V(\bbr)\right)=\det \hat{Y}(L).
\end{eqnarray}
Similarly to the 1D case, this formula should be either properly regularised or considered on a discrete lattice.
This result enables one to reduce calculation of the $d$-dimensional functional determinant to the computation of the $(d-1)$-dimensional determinant, which can drastically simplify the problem. Even when a solution of the initial value problem is not known explicitly, the above relation can be useful for developing some approximations. As it is explained below, it is also very convenient for numerical calculation of functional determinants. In this work, we apply the generalised GY formula to the massive and massless two-dimensional Laplace operators, which allows us to derive asymptotic expressions for their determinants in the discrete case.

The paper is organised as follows. In Section \ref{1D-case-section} we derive the one-dimensional GY formula for the discrete and continuum case by a method, which can be easily generalised to higher dimensions. In Section \ref{d-dim-section} such a generalisation is obtained. Finally, in Section \ref{section-laplace2d} we apply our generalised GY formula to the two-dimensional massive and massless Laplace operators and find asymptotic expressions for their determinants.  


\section{One-dimensional case}\label{1D-case-section}

In this section we reproduce the Gelfand-Yaglom (GY) formula in the one-dimensional case.


\subsection{Recursion relations for the determinant in the discrete model}

Our approach is similar to the original method by Gelfand and Yaglom \cite{GY60}, but in order to get the recursion relations for the determinants we represent the determinants in terms of the Gaussian integrals. This enables us to generalise easily our approach to the higher dimensional case. Consider the discrete one-dimensional Hamiltonian
\be\label{Ham1d-def}
(H\psi)_i=2\psi_i-\psi_{i+1}-\psi_{i-1}+V_i\psi_i,\quad i=1,\dots, N-1,
\en
which is the sum of the Laplace operator and the potential term and we assume the Dirichlet boundary conditions: $\psi_0=\psi_N=0$.

The determinant of this operator can be calculated with help of the Gaussian integrals, which can be evaluated recursively using a set of the functions $\Phi_n(x)$ defined recursively as
\be\label{phi_recursion}
\Phi_{n+1}(x)=e^{-V_{n+1}x^2}\intinf \frac{dy}{\sqrt{\pi}}\; e^{-(x-y)^2}\Phi_n(y).
\en
It is easy to check that
\be\label{det-phi}
\left(\det(H)\right)^{-\frac{1}{2}}=\Phi_{N}(0),
\en
provided that $\Phi_1(x)=e^{- V_{1}x^2- x^2}$.

Since the initial function is Gaussian, all $\Phi_n(x)$ are Gaussian as well, so we can assume that
\be
\Phi_n(x)=c_ne^{-a_n x^2-b_nx}.
\en
Substituting this ansatz into Eq.(\ref{phi_recursion}) we obtain the following set of the recursion relations:
\begin{eqnarray}
\label{a-rec-rel} a_{n+1}&=&V_{n+1} +1-\frac{1}{a_n+1},\quad a_0=\infty,\\
b_{n+1}&=&\frac{b_n}{a_n+1},\quad b_1=0,\\
c_{n+1}&=&\frac{c_n}{\sqrt{(a_n+1)}}\:e^{\frac{b_n^2}{4(a_n+1)}}, \quad c_1=1.
\end{eqnarray}
As $b_1=0$ it follows from the second relation that $b_n=0$ and 
\begin{eqnarray}
c_{n+1}=\prod_{k=1}^n \frac{1}{\sqrt{(a_k+1)}}.
\end{eqnarray}

Then the formula for the determinant in terms of $a_n$  reads
\begin{eqnarray}
\label{det-rec-rel}\det H &=&\frac{1}{c_N^2}=\prod_{k=1}^{N-1}(a_k+1).
\end{eqnarray}

This formula relates the determinant to coefficients $a_n$, which can be computed recursively for an arbitrary potential via Eq.\eqref{a-rec-rel}.

Introducing new variables $y_n$ such that 
\begin{eqnarray}\label{y-def}
a_n+1=\frac{y_{n+1}}{y_n}, \quad y_0=0,\; y_1=1.
\end{eqnarray}
we can rewrite the recursion relation for $a_n$ as
\begin{eqnarray}
-(y_{n+2}-2y_{n+1}+y_n)+V_{n+1}y_{n+1}=0.
\end{eqnarray}
Thus $y_n$ is a solution of the equation $(Hy)_n=0$, satisfying the initial condition $y_0=0,\; y_1=1$. The expression for the determinant \eqref{det-rec-rel} in terms of $y_n$ reads
\begin{eqnarray}\label{det-y}
\det H&=&\prod_{k=1}^{N-1}(a_k+1)=\prod_{k=1}^{N-1}\frac{y_{k+1}}{y_k}=y_N.
\end{eqnarray}
This is a standard discrete form of the one-dimensional GY formula \cite{D12, F92}.


\subsection{GY formula in the continuum model}

In order to rewrite Eq.\eqref{a-rec-rel} in the continuum model we set $n=x/\dell$,  $V_{n+1}=\dell \int_x^{x+\dell} dx' V(x')$, $a_n=z_n\dell=z(n\dell)\dell= z(x) \dell$. Substituting these expressions into Eq.\eqref{a-rec-rel} and keeping only the leading terms in $\dell$ one finds
\begin{eqnarray}
\frac{z_{n+1}-z_n}{\dell}&=&-z_n^2+\frac{1}{\dell}\int_x^{x+\dell} dx' V(x')+O(\dell).
\end{eqnarray}
Taking the limit $\dell \to 0$, $N\to \infty$, such that $L=N\dell$ is a constant, one obtains
\begin{eqnarray}\label{z-diff-eqn}
\frac{dz}{dx}=-z(x)^2+V(x), \quad z(0)=\infty.
\end{eqnarray}
Eq.\eqref{det-rec-rel} is then transformed into
\begin{eqnarray}\label{mod-det-1d}
|\det H| &=&\det \left(-\frac{d^2}{d x^2}+V(x)\right) =\lim_{\dell\to 0}\prod_{k=1}^{N-1}(1+z_k\dell)=e^{\int_0^Ldx z(x)}.
\end{eqnarray}
The appearance of the absolute value on the left hand side is due to the fact that if $\det H$ changes its sign, then $z_n$ becomes of order of $1/\dell$ for some $n$ and contributions of such terms, which must be treated separately, compensate the overall sign of the determinant. This is the first version of the one-dimensional GY formula in the continuum, expressing $\det H$ through the solution of the first-order non-linear Riccati differential equation.

In order to get another formulation of the GY formula we take the continuum limit of Eq.\eqref{det-y} and find
\begin{eqnarray}
-y''+Vy=0,\quad y(0)=0,\; y'(0)=1, \quad \det \left(-\frac{d^2}{d x^2}+V(x)\right)= y(L).
\end{eqnarray}
Using this result and the substitution $z(x)=\frac{y'(x)}{y(x)}$ one can immediately recover \eqref{mod-det-1d}, as $\int dx z(x)=\ln |y|$. This is the standard GY formula. As the determinant diverges in the limit $N\to \infty$, it should be regularised in some way. 


\section{$d$-dimensional case}\label{d-dim-section}

\subsection{Recursion relations for the determinant in the discrete model}

We consider the discrete  Hamiltonian 
\begin{eqnarray}
(H\psi)_{\boldsymbol{i}}=-(\Delta_d \psi)_{\boldsymbol{i}}+v_{\boldsymbol{i}}\psi_{\boldsymbol{i}},
\end{eqnarray}
defined on a $d$-dimensional cubic lattice of the dimension $(N-1)\times (M-1)^{d-1}$, where $\Delta_d$ is the $d$-dimensional discrete Laplacian and the Dirichlet boundary condition is imposed in all directions. It is convenient to separate the transverse and longitudinal directions in the Hamiltonian by introducing the vectors $\bpsi_i=(\psi_{i1},\psi_{i2},\dots,\psi_{iK})^{T}$ and the diagonal matrices $V_i=\diag(v_{i1},v_{i2},\dots,v_{iK})$ with $K\equiv (M-1)^{d-1}$, then $H$ can be represented as
\be\label{Ham2d-def}
(H\bpsi)_i=2\bpsi_i-\bpsi_{i+1}-\bpsi_{i-1}+(V_i-\Laplayer)\bpsi_i,\quad i=1,\dots, N-1,
\en
where $\Laplayer$ is the $(d-1)$-dimensional discrete Laplacian. 

Similarly to the one-dimensional case, we introduce a set of the functions $\Phi_n(\bx)$ defined recursively as
\be\label{phi_recursion-d}
\Phi_{n+1}(\bx)=e^{-(\bx,(-\Laplayer+V_{n+1}) \bx)}\intinf d\by\; e^{-(\bx-\by)^2}\Phi_n(\by),
\en
where $\bx=(x_{1},x_{2},\dots,x_{K})^T$, $d\by=\prod_{k=1}^{K}\frac{dy_{k}}{\sqrt{\pi}}$ and $(\bx,\by)$ denotes the standard scalar product in $\mathbb{R}^K$. Then the  determinant of $H$ is given by 
\be\label{det-phi-d}
\left(\det H\right)^{-\frac{1}{2}}=\Phi_{N}(0),
\en
provided that $\Phi_1(\bx)=e^{-(\bx,(-\Laplayer+ V_{1}+I)\bx)}$.

Since the initial function is Gaussian,  $\Phi_n(\bx)$ must be again Gaussian: 
\be
\Phi_n(\bx)=c_ne^{-(\bx,A_n \bx)-(\bx,\bb_n)}
\en
Substituting this ansatz into Eq.(\ref{phi_recursion-d}) we obtain the following set of the recursion relations:
\begin{eqnarray}\label{a-rec-rel-d}
A_{n+1}&=&I-(A_n+I)^{-1}-\Laplayer+V_{n+1}, \quad A_0=\infty,\\ 
\bb_{n+1}&=&(A_n+I)^{-1}\bb_n,\quad \bb_1=0,\\
c_{n+1}&=&\frac{c_n}{\sqrt{\det(A_n+I)}}\:e^{\frac{1}{4}(\bb_n,(A_n+I)^{-1}\bb_n)},\quad c_1=1.
\end{eqnarray}
As $\bb_1=0$ it follows from the second relation that $\bb_n=0$ and
\begin{eqnarray}
c_{n+1}=\prod_{k=1}^n\frac{1}{\sqrt{\det(A_k+I)}}.
\end{eqnarray}
Then the formula for the determinant in terms of the matrices $A_n$, which can be computed recursively for an arbitrary potential via Eq.\eqref{a-rec-rel-d}, reads
\begin{eqnarray}
\label{det-rec-rel-d}\det H &=&\frac{1}{c_N^2}=\prod_{k=1}^{N-1}\det(A_k+I).
\end{eqnarray}

Introducing new set of matrices $Y_n$ such that 
\begin{eqnarray}
A_n+I=Y_{n+1}Y_n^{-1}, \quad Y_0=0,\; Y_1=I.
\end{eqnarray}
we can rewrite the recursion relation for $A_n$ as
\begin{eqnarray}
\label{GY-disc-Y}&&-(Y_{n+2}-2Y_{n+1}+Y_n)+(-\Laplayer+V_{n+1})Y_{n+1}=0.
\end{eqnarray}
Thus $Y_n$ is a matrix solution of the equation $(HY)_n=0$, satisfying the initial condition $Y_0=0,\; Y_1=I$. The expression for the determinant \eqref{det-rec-rel-d} in terms of $Y_n$ reads
\begin{eqnarray}\label{det-y-d}
\det H&=&\prod_{k=1}^{N-1}\det(A_k+I)=\prod_{k=1}^{N-1}\det Y_{k+1}Y_k^{-1}=\det Y_N.
\end{eqnarray}
This is a discrete form of the $d$-dimensional GY formula. 

We note that one of the possible applications of this formula is that it provides an efficient numerical algorithm for computing the functional determinants. Indeed, the original matrix of $H$ has the dimension $N^d\times N^d$, for $N=M$, and hence the running time of numerical algorithms calculating its determinant directly scales as $N^{3d}$. At the same time, the matrices involved in the recursion relations \eqref{a-rec-rel-d} or \eqref{GY-disc-Y} have the dimension $N^{d-1}\times N^{d-1}$ and therefore the time needed to compute the determinant using Eq.\eqref{det-y-d} scales as $N^{3d-2}$.

\subsection{Recursion relations in the continuum model}


The matrices in the discrete formulation correspond to the position representation, which is not very convenient if we want to take the continuum limit. Therefore we can first rewrite all formulas above in any other representation, which admits a countable basis $\{\ket{u_k}\}$ in the continuum limit. Any operator $C$ in this representation is represented  by a matrix $\hat{C}$, such that  $\hat{C}_{kl}=\bbra{u_k}C\ket{u_l}$. For example, Eq.\eqref{a-rec-rel-d} takes the form
\begin{eqnarray}\label{hatA-rec-rel}
\hat{A}_{n+1}&=&I-(\hat{A}_n+I)^{-1}-\hat{\Delta}_{d-1}+\hat{V}_{n+1}, \quad \hat{A}_0=\infty.
\end{eqnarray}

The most natural choice for the basis functions in the transverse direction is given by the orthonormal eigenfunctions $u_{\bn}(\bbrho)$, $\bbrho=(x_1,\dots, x_{d-1})$, $\bn=(n_1,\dots n_{d-1})$ of $\Laplayer$:
\begin{eqnarray}\label{u-def}
u_{\bn}(\bbrho)=\left(\frac{2}{W}\right)^{\frac{d-1}{2}}\prodrho\sin \frac{\pi n_i x_i}{W},\quad n_i\in \mathbb{N}.
\end{eqnarray}
Then the matrix elements of $\hat{V}(x)$ are given by 
\begin{eqnarray}\label{Vmatrix-cont}
\left(\hat{V}(x)\right)_{\bn \bm}=\int d\bbrho V(x,\bbrho)u_{\bn}(\bbrho)u_{\bm}(\bbrho).
\end{eqnarray}

In order rewrite Eq.\eqref{hatA-rec-rel} in the continuum model we set $n=x/\dell$, $\hat{\Delta}_{d-1}=\dell^2\hat{\Delta}_{d-1}^{{\rm cont}}$, where $\Delta_{d-1}^{{\rm cont}}$ stands for the continuum Laplacian, $\hat{V}_{n+1}=\dell \int_x^{x+\dell} dx' \hat{V}(x')$ and $\hat{A}_n=\hat{Z}_n\dell=\hat{Z}(n\dell)\dell= \hat{Z}(x) \dell$. Then the recursion relation \eqref{hatA-rec-rel} is transformed into
\begin{eqnarray}
\frac{\hat{Z}_{n+1}-\hat{Z}_n}{\dell}&=&-\hat{\Delta}_{d-1}^{{\rm cont}}-\hat{Z}_n^2+\frac{1}{\dell} \int_x^{x+\dell} dx' \hat{V}(x')+O(\dell).
\end{eqnarray}
Taking the limit $\dell \to 0$, $N, M\to \infty$ and keeping the products $L=N\dell$ and $W=M\dell$ finite, one obtains
\begin{eqnarray}\label{z-diff-eqn-d}
\frac{d\hat{Z}}{dx}=-\hat{\Delta}_{d-1}-\hat{Z}(x)^2+\hat{V}(x), \quad \hat{Z}(0)=\infty,
\end{eqnarray}
where we omitted the superscript "cont" for the Laplacian. This equation generalises the one-dimensional result Eq.\eqref{z-diff-eqn}.

The formula \eqref{det-rec-rel-d} for the determinant is transformed into the following form:
\begin{eqnarray}\label{det-z-cont-d}
|\det H| &=&\det \left(-\Delta_d+V(\bbr)\right) =\lim_{\dell\to 0}\prod_{k=1}^{N-1}\det(1+\hat{Z}_k\dell)=
\e^{\tr\int_0^Ldx  \hat{Z}(x)}.
\end{eqnarray}
Thus we derived a relation between the functional determinant of $H$ and the solution of the matrix Riccati differential equation.

Similarly to the 1D case, the second formulation of the GY formula can be obtained by taking the continuum limit of Eq.\eqref{GY-disc-Y} 
\begin{eqnarray}\label{eqn-Y-d}
-\hat{Y}''+(-\hat{\Delta}_{d-1}+\hat{V})\hat{Y}=0, \quad \hat{Y}(0)=0,\;\hat{Y}'=I,
\end{eqnarray}
and the formula for the determinant reads
\begin{eqnarray}\label{GY-cont-Y-d}
\det H &=&\det \left(-\Delta_d+V(\bbr)\right)=\det \hat{Y}(L).
\end{eqnarray}
Again, Eq.\eqref{det-z-cont-d} follows from this formula and the relation $\hat{Z}=\hat{Y}'\hat{Y}^{-1}$.
This result represents the $d$-dimensional generalisation of the GY formula. The determinants in the continuum limit should be properly regularised. One possible way to do it is to calculate them first in the discrete formulation and then analyse the asymptotic behaviour of the obtained result in the continuum limit. It is known, for example, that for the two-dimensional massless Laplacian the result for the $\zeta$-regularised determinant can be extracted from the corresponding asymptotic expression for the discrete Laplacian \cite{DD88}. 

Equations \eqref{z-diff-eqn-d},  \eqref{det-z-cont-d}, \eqref{eqn-Y-d}, \eqref{GY-cont-Y-d} and their discrete counterparts represent the main result of this work.


\section{Example: determinants of the two-dimensional massive and massless Laplace operators}\label{section-laplace2d}

Consider the determinant of the two-dimensional massive Laplace operator $H=-\Delta_2+m^2$ corresponding to $V(\bbr)=m^2$. Eq.\eqref{GY-disc-Y} then reads
\begin{eqnarray}
&&Y_{n+2}+Y_n+(\Delta_1-2-m^2)Y_n=0,\quad Y_0=0,\; Y_1=I,\\
&& (\Delta_1 \psi)_i=\psi_{i+1}+\psi_{i-1}-2\psi_i.
\end{eqnarray}
Since the determinant of $H$ is equal to the determinant of $Y_N$, we can compute $Y_N$ in a any basis, in particular, we can choose the eigenbasis of $\Delta_1$. Then $\Delta_1=\diag (\{\lambda_k\})$, with $\lambda_k=-2(1-\cos \frac{\pi k}{M}),\:k=1,\dots, M-1$, and $Y_n$ is also diagonal $Y_n\equiv\diag (\{\mu^{(n)}_k\})$ and  $\mu^{(n)}_k$ satisfy the relation
\begin{eqnarray}\label{mu_k-eqn}
&&\mu^{(n+2)}_k+\mu^{(n)}_k-(m^2+2-\lambda_k)\mu^{(n)}_k=0,\nonumber\\
&& \mu^{(0)}_k=0,\; \mu^{(1)}_k=1,\quad k=1,\dots, M-1
\end{eqnarray}
Once $\mu^{(n)}_k$ are found we obtain the result for $\det H$:
\begin{eqnarray}
\det H= \prod_{k=1}^{M-1}\mu^{(N)}_k.
\end{eqnarray}
The solution of Eq.\eqref{mu_k-eqn} can be found as a linear combination of the two solutions $e^{\pm \gamma_k n}$, where $\gamma_k$ is determined by the condition
\begin{eqnarray}\label{gamma-eqn}
&&e^{\gamma_k}+e^{-\gamma_k}-(m^2+2-\lambda_k)=0,\\
&&\gamma_k=\arccosh \left(1+\frac{m^2-\lambda_k}{2}\right).
\end{eqnarray}
The solution satisfying the initial condition is given by
\begin{eqnarray}
\mu^{(n)}_k=\frac{\sinh \gamma_k n}{\sinh \gamma_k},
\end{eqnarray}
and the expression for the determinant reads
\begin{eqnarray}
\det (-\Delta_2+m^2)=\prod_{k=1}^{M-1}\frac{\sinh \gamma_k N}{\sinh \gamma_k}.
\end{eqnarray}


\subsection{Asymptotic behaviour of the determinant of the two-dimensional massive Laplace operator}\label{subsection-asymptotic}

In this section we calculate the asymptotic behaviour of the above result for the determinant for $N, M\to \infty$, at a fixed value of $N/M$. To this end we first rewrite the $\frac{\sinh \gamma_k N}{\sinh \gamma_k}$ as follows
\begin{eqnarray}
\frac{\sinh \gamma_k N}{\sinh \gamma_k}=\frac{e^{\gamma_k N}-e^{-\gamma_k N}}{e^{\gamma_k }-e^{-\gamma_k}}=
\xi_k^{N}\frac{1-\xi_k^{-2N}}{\xi_k-\xi_k^{-1}}, \quad \xi_k\equiv e^{\gamma_k}.
\end{eqnarray}
Then we find
\begin{eqnarray}\label{S1-S2-S3}
&&\ln \det H=S_1+S_2+S_3,\quad S_1\equiv N\sum_{k=1}^{M-1}\ln \xi_k,\;  S_2\equiv\sum_{k=1}^{M-1}\ln (1-\xi_k^{-2N}),\nonumber\\
&& S_3\equiv-\sum_{k=1}^{M-1}\ln (\xi_k-\xi_k^{-1}).
\end{eqnarray}
In order to evaluate these sums we write $\sum_{k=1}^{M-1}f(k)=\sum_{k=0}^{M}f(k)-(f(0)+f(M))$ and  apply the Euler-MacLaurin formula:
\be\label{Euler-MacLaurin}
\sum_{k=1}^{M-1}f(k)=\int_0^M dk f(k)-\frac{1}{2}\left(f(M)+f(0)\right) +
\sum_{n=1}^{\infty}\frac{B_{2n}}{(2n)!}\left(f^{(2n-1)}(M)-f^{(2n-1)}(0)\right),
\en
where $B_{2n}$ are the Bernoulli numbers, $B_2=1/6$, $B_4=-1/30$.

From Eq.\eqref{gamma-eqn} one can find an explicit expression for $\xi_k$:
\begin{eqnarray}
\xi_k=1+\frac{m^2-\lambda_k+\sqrt{(m^2-\lambda_k)^2+4(m^2-\lambda_k)}}{2},
\end{eqnarray}
and therefore
\begin{eqnarray}
\lambda_0&=&0, \quad \xi_0=1+\frac{m^2+\sqrt{m^4+4m^2}}{2},\nonumber\\
\lambda_M&=&-4, \quad \xi_M=1+\frac{m^2+4+\sqrt{(m^2+4)^2+4(m^2+4)}}{2}.
\end{eqnarray}

Thus for the first sum $f(0)=\ln\left(1+\frac{m^2+\sqrt{m^4+4m^2}}{2}\right)=\arccosh \left(1+\frac{m^2}{2}\right)$, $f(M)=\ln\left(1+\frac{m^2+4+\sqrt{(m^2+4)^2+4(m^2+4)}}{2}\right)=\arccosh \left(3+\frac{m^2}{2}\right)$ and $f'(0)=f'(M)=0$ (this is true for $m\neq 0$, the massless case $m=0$ will be considered separately below). For the higher order derivatives we get $f''(0)=O(M^{-2})$ and $f''(M)=O(M^{-2})$ and such terms will be neglected.

The integral $\int_0^M dk f(k)=\frac{M}{\pi}\int_0^{\pi} dx f\left(\frac{Mx}{\pi}\right)\equiv M I_1(m)$ with 
\begin{eqnarray}
&&I_1(m)\equiv \frac{1}{\pi}\int_0^\pi dx \ln  \phi(x)
= \frac{1}{\pi}\int_0^\pi dx \arccosh \left(1 + \frac{m^2 + 2 (1 - \cos x)}{2}\right),\nonumber\\
\end{eqnarray}
where $\phi(x)\equiv 1+\frac{m^2+2(1-\cos x)+\sqrt{(m^2+2(1-\cos x))^2+4(m^2+2(1-\cos x))}}{2}$, cannot be calculated analytically. The expression for $S_1$ reads
\begin{eqnarray}
S_1= NM I_1(m)-\frac{N}{2}g(m)+O\left(\frac{N}{M^{2}}\right),
\end{eqnarray}
with $g(m)\equiv \arccosh \left(1+\frac{m^2}{2}\right)+\arccosh \left(3+\frac{m^2}{2}\right)$.

The second sum $S_2$ converges exponentially fast to its limiting value, which is determined by the behaviour of $\xi_k$ at small $k$:
\begin{eqnarray}
\lim_{M\to \infty}S_2&=&\lim_{M\to \infty}\sum_{k=1}^{M-1}\ln\left(1-
\left(c_0(m)+c_1(m)\frac{k^2}{2 M^2}+O\left(\frac{k^4}{M^4}\right)\right)^{-2N}\right)\nonumber\\
&=&\lim_{M\to \infty}\sum_{k=1}^{M-1}\ln\left(1-c_0(m)^{-2N}
\exp{-\frac{c_1(m)N k^2}{c_0(m)M^2}+O\left(\frac{k^4N}{M^4}\right)}\right)\nonumber\\
&=&\sum_{k=1}^{\infty}\ln\left(1-c_0(m)^{-2N} e^{-\frac{c_1(m)N k^2}{c_0(m)M^2}}\right),
\end{eqnarray}
where $c_0(m)=\frac{1}{2}\left(m^2+2+\sqrt{m^2(m^2+4)}\right)$, $c_1(m)=\frac{\pi^2}{2}\left(1+\frac{m^2+2}{\sqrt{m^2(m^2+4)}}\right)$. Since this sum is exponentially small in $N$ it can be neglected.

For the last term $S_3$, we can again apply the  Euler-MacLaurin formula with $f(k)=-\ln (\xi_k-\xi_k^{-1})=-\frac{1}{2}\ln \left(m^4+8m^2+14 -4(m^2+4)\cos \left(\frac{\pi k}{M}\right)+2\cos \left(\frac{2\pi k}{M} \right)\right)$. We find $f(0)=-\frac{1}{2}\ln \left(m^2(m^2+4)\right)$, $f(M)=-\frac{1}{2}\ln \left((m^2+4)(m^2+8)\right)$, $f'(0)=f'(M)=0$, $f''(0)=O(M^{-2})$, $f''(M)=O(M^{-2})$ and the integral term yields $MI_2(m)$ with 
\begin{eqnarray}
I_2(m)\equiv -\frac{1}{2\pi}\int_0^{\pi}dx \ln \left(m^4+8m^2+14 -4(m^2+4)\cos x+2\cos 2x \right).
\end{eqnarray}
Therefore we get
\begin{eqnarray}
S_3=MI_2(m)+\frac{1}{4}\ln \left(m^2(m^2+4)^2(m^2+8)\right)+O\left(NM^{-2}\right).
\end{eqnarray}
Collecting the results for $S_1$, $S_2$ and $S_3$ we obtain
\begin{eqnarray}
\ln \det (-\Delta_2+m^2)&=& NM I_1(m)-\frac{N}{2}g(m)+MI_2(m)\nonumber\\
&+&\frac{1}{4}\ln \left(m^2(m^2+4)^2(m^2+8)\right)+O\left(\frac{N}{M^{2}}\right).
\end{eqnarray}
The above expression must be invariant under the exchange of $N$ and $M$ and one can check that $I_2(m)=-\frac{1}{2}\left(\arccosh \left(1+\frac{m^2}{2}\right)+\arccosh \left(3+\frac{m^2}{2}\right)\right)$ and hence
\begin{eqnarray}
\ln \det (-\Delta_2+m^2)&=& NM I_1(m)-\frac{N+M}{2}g(m)\nonumber\\
&+&\frac{1}{4}\ln \left(m^2(m^2+4)^2(m^2+8)\right)+O\left(\frac{N}{M^{2}}\right).
\end{eqnarray}


\subsection{Massless Laplacian}

The same strategy can be used in the massless case, $m=0$, however more terms in the Euler-MacLaurin formula give contributions to the final result and some of them should be treated more carefully. 

Our starting point Eq.\eqref{S1-S2-S3} is the same as in the previous section. Applying the Euler-MacLaurin formula \eqref{Euler-MacLaurin} to the first sum and taking into account  that $f(0)=0$, $f(M)=\ln(3+2\sqrt{2})$, $f'(0)=\pi/M$, $f'(M)=0$  one obtains
\begin{eqnarray}
S_1&=&\frac{NM}{\pi}\int_0^{\pi}dx \ln h(x)-\frac{1}{2}N\ln(3+2\sqrt{2})-\frac{\pi}{12}\frac{N}{M}+O\left(\frac{N}{M^{2}}\right) \nonumber\\
&=& \frac{4G}{\pi}NM-N\ln(1+\sqrt{2})-\frac{\pi}{12}\frac{N}{M}+O\left(\frac{N}{M^{2}}\right),
\end{eqnarray}
where $h(x)\equiv 2-\cos x+\sqrt{3-4\cos x+\cos^2x}$ and  $G$ is the Catalan constant. 

The second sum $S_2$ converges exponentially fast to its limiting value, which is again determined by the behaviour of $\xi_k$ at small $k$:
\begin{eqnarray}
\lim_{M\to \infty}S_2&=&\lim_{M\to \infty}\sum_{k=1}^{M-1}\ln\left(1-\left(1+\frac{\pi k}{M}+O\left(\frac{k^2}{M^2}\right)\right)^{-2N}\right)\nonumber\\
&=&\sum_{k=1}^{\infty}\ln\left(1-e^{-\frac{2\pi k N}{M}}\right)= \ln P(q),
\end{eqnarray}
where $q=e^{-2\pi N/M}$ and $P(q)=\prod_{k=1}^{\infty}(1-q^k)$.

A direct application of the Euler-MacLaurin formula to $S_3$ is impossible due to the divergent derivative of the function $f(k)$ at $k=0$. In order to over come this problem we may write
\be
S_3=-\sum_{k=1}^{M-1}\left(\ln\left(\xi_k-\xi_k^{-1}\right)-\ln\frac{\pi k}{M}\right)-\sum_{k=1}^{M-1}\ln\frac{\pi k}{M}\equiv 
S_3^{(1)}+S_3^{(2)}.
\en 
The first sum in the above equation can be now calculated with the help the Euler-MacLaurin formula with $f(0)=\ln 2$, $f(M)=\ln (4\sqrt{2}/\pi)$, $f'(0)=0$, $f'(M)=-1/M$:
\begin{eqnarray}
S_3^{(1)}&=&-\frac{M}{\pi}\int_0^{\pi}dx \left(\ln j(x)-\ln x\right)
+\frac{1}{2}\ln\left(\frac{8\sqrt{2}}{\pi}\right)+\frac{1}{12M} +\dots \nonumber\\
&=&-M\left(\ln(1+\sqrt{2})-\ln\pi +1\right)+\frac{1}{2}\ln\left(\frac{8\sqrt{2}}{\pi}\right)+\frac{1}{12M} +\dots,
\end{eqnarray}
where $j(x)\equiv 2\sqrt{3-4\cos x+\cos^2x}$ and dots denote the terms vanishing in the limit $N, M\to \infty$.
The last sum $S_3^{(2)}$ can be evaluated using the Stirling's formula:
\begin{eqnarray}
S_3^{(1)}&=&-\sum_{k=1}^{M}\ln\frac{\pi k}{M}+\ln\pi=-M(\ln\pi-\ln M)+\ln\pi-\ln M!\nonumber\\
&=&M\ln M-(M-1)\ln\pi-
\ln\left(\sqrt{2\pi M}\left(\frac{M}{e}\right)^M\left(1+\frac{1}{12M}+\dots\right)\right)\nonumber\\
&=&-M(\ln\pi-1)-\frac{1}{2}\ln M-\frac{1}{2}\ln\left(\frac{2}{\pi}\right)-\frac{1}{12M} +\dots.
\end{eqnarray}
Thus we find for $S_3$
\be
S_3=-M\ln(1+\sqrt{2})-\frac{1}{2}\ln M+\frac{1}{2}\ln(4\sqrt{2})+O\left(\frac{1}{M^2}\right).
\en
Adding all the results we obtain
\begin{eqnarray}
\ln\left(\det \Delta_2\right)
&=&\frac{4G}{\pi}NM-(N+M)\ln(1+\sqrt{2})-\frac{1}{4}\ln (NM)\nonumber\\
&&+\frac{1}{2}\ln(4\sqrt{2})+\ln\left(q^{\frac{1}{24}}\left(\frac{N}{M}\right)^{\frac{1}{4}} P(q)\right)+O\left(\frac{N}{M^{2}}\right).
\end{eqnarray}
One can show that the expression in the last logarithm is invariant under exchange of $N$ and $M$ \cite{DD88}, thus making the whole result to be invariant as well. This formula is in agreement with Eq.(4.20) of Ref.\cite{DD88}, where the determinant of the two-dimensional massless Laplace operator was calculated by a different method. 


\section{Conclusions}

We have derived a formula for the determinant of the $d$-dimensional operator $-\Delta +V(\bbr)$, which generalises the Gelfand-Yaglom result in one dimension. The derivation is based on the Gaussian integral representation of the determinant, which allows us to derive the recursion relations similar to those found by Gelfand and Yaglom in the one-dimensional case. It is remarkable, that they have exactly the same structure as in $1D$, in spite of the non-commutativity of the matrices involved in the relations. Similar results involving non-commutative matrices were derived before for one-dimensional matrix-valued ordinary differential operators (see \cite{Kleinert_book, FFG17jpa} and references therein). Our derivation was applied first to the discrete case, when the operators are defined on a lattice, and then the corresponding formula in the continuum was obtained in the limit when the lattice constants goes to zero, while the number of lattice points tends to infinity. We note that our approach does not require any symmetries of the potential term, in contrast to previous attempts to generalise the GY formula to higher dimensions \cite{DK06}.

Similarly to the $1D$ case, we expect that the solution of the corresponding initial value problem can be found explicitly only for very special types of potential. However, we believe that our formula can be generally useful for an approximate calculation of the determinant, provided that there is a small or large parameter in the Hamiltonian. For example, for $H=-\Delta +V(\bbr)-E$, such a small parameter might be $|V(\bbr)/E|$, which is a typical situation in the field of  disordered systems \cite{HJ91,FF15,FFG17}. Our method also provides a very efficient way of calculating the functional determinants numerically. 

In order to illustrate our main result, we applied the generalised GY formula to the determinants of the two-dimensional massive and massles Laplace operators on a rectangular domain. For the massless case we reproduced the well known result derived previously by a different method \cite{DD88}. For the massive Laplace operator on a sphere, the results are available in the literature \cite{D94, D14, CEZ15}, however we are not aware of the corresponding results for a rectangular domain.

In the present work we considered only the Dirichlet boundary condition and assumed that the differential operator has no zero modes. It would be interesting to develop a similar approach to other types of the boundary conditions and extend it to the situation where zero modes are present \cite{F87, MT95, FFG17jpa, KC99}.

\subsection{Acknowledgments} I am grateful to Yan Fyodorov for critical reading of the manuscript and useful comments.


\end{document}